\documentclass[12pt]{elsart}

\usepackage{pstricks}
\usepackage{amssymb,amsmath}
\newpsobject{grilla}{psgrid}
{subgriddiv=1,griddots=10,gridlabels=6pt}
\newcommand{\gsim}
{\mbox{${~\raise.25em\hbox{$>$}\kern-.70em
\lower.25em\hbox{$\sim$}~}$}}
\newcommand{\lsim}
{\mbox{${~\raise.25em\hbox{$<$}\kern-.70em
\lower.25em\hbox{$\sim$}~}$}}
\newcommand{\beq}{\begin{equation}}
\newcommand{\beqa}{\begin{eqnarray}}
\newcommand{\eeq}{\end{equation}}
\newcommand{\eeqa}{\end{eqnarray}}
\newcommand{\AddrAHEP}
{{\it AHEP Group, Instituto de F\'{\i}sica Corpuscular -
    CSIC/Universitat de Val{\`e}ncia \\
    Edificio de Institutos de Paterna, Apartado 22085,
  E--46071 Val{\`e}ncia, Spain}}
\begin{document}
\begin{frontmatter}
  \begin{flushright}
    IFIC/07-24\\
    PI/UAN-FT2007-172\\
  UdeA-PE-07/004  
  \end{flushright}
\title{Variations on leptogenesis
}
\author[ahep]{Diego Aristizabal Sierra}
\author[narino]{Marta Losada}
\author[frascati,udea]{Enrico Nardi}
\address[ahep]{\AddrAHEP}
\address[narino]{{\normalsize \it 
Centro de Investigaciones, Universidad Antonio Nari\~{n}o, \\
Cra 3 Este No 47A-15 Bloque 4, Bogot\'{a}, Colombia}}
\address[frascati]{{\normalsize \it INFN, 
    Laboratori Nazionali di Frascati, 
    C.P. 13, I00044 Frascati, Italy}}
\address[udea]{{\normalsize\it 
Instituto~de~F\'\i sica,~Universidad~de~Antioquia,~A.A.{\it{1226}},~Medell\'\i n,~Colombia}}
\begin{abstract}
  \noindent
  We study variations of the standard leptogenesis scenario that can arise if
  an additional mass scale related to the breaking of some new symmetry (as
  for example a flavor or the B-L symmetry) is present below the mass
  $M_{N_1}$ of the lightest right-handed Majorana neutrino.  Our
  scheme is inspired by $U(1)$ models of flavor \`a la Froggatt-Nielsen, and
  involves new vectorlike heavy fields $F$.  We show that depending on the
  specific hierarchy between $M_{N_1}$ and the mass scale of the fields $F$,
  qualitatively different realizations of leptogenesis can emerge. We compute
  the $CP$ asymmetries in $N_1$ decays in all the relevant cases, and we
  conclude that in most situations leptogenesis could be viable at scales much
  lower than in the standard scenario.
\end{abstract}

\end{frontmatter}

\section{Introduction}
\label{sec:introduction}

\vspace{-.2cm} 
Baryogenesis through leptogenesis represents an attractive mechanism to
explain the observed matter-antimatter asymmetry of the Universe
\cite{fu86,lu92}.  Once the Standard Model (SM) is extended by including the
see-saw mechanism in order to explain the strong suppression of the neutrino
mass scale \cite{seesaw}, the possibility of generating a cosmic lepton
asymmetry via the lepton number and $CP$ violating out-of-equilibrium decays
of the seesaw singlet neutrinos arises as a natural possibility.  Partial
conversion of the lepton asymmetry into a baryon asymmetry then proceeds by
means of electroweak sphaleron interactions \cite{kuz85} that are
non-perturbative SM processes.  Qualitatively, it is almost
unavoidable that a model that includes the seesaw mechanism will predict a
certain amount of matter-antimatter asymmetry surviving until the present
epoch, and then the question of whether leptogenesis is able to explain the
puzzle of the baryon asymmetry of the Universe is essentially a quantitative
one.  In recent years, quantitative analysis of the standard leptogenesis
scenario have become more and more sophisticated, taking into account many
subtle but significant ingredients, such as various washout effects
\cite{BBP0204-BP9900,pi04-pi05,ha04}, thermal corrections to particle masses
and $CP$ violating asymmetries \cite{gi04}, spectator processes
\cite{bu01-na05}, flavor effects
\cite{ba00,en03-fu05,aba06a-aba06b,na06,DeSimone:2006dd,Josse-Michaux:2007zj,Shindou:2007se}
and the possible effects of the heaviest right handed Majorana neutrinos
$N_{2,3}$ \cite{ba00,diba05,vi05,eng06} (for reviews of the most recent
results see \cite{review}).

One assumption that is common to all these studies is that between the scale
of the breaking of lepton number and the electroweak breaking scale, there are
no additional sources of new physics that could affect the mechanism of
leptogenesis. This assumption is certainly justified both in terms of
simplicity and also because it allows for a certain level of predictivity,
that is mainly due to the fact that the same couplings that determine the $CP$
asymmetries and the out-of-equilibrium conditions in the decays of the heavy
Majorana singlets are also responsible for the seesaw masses of the light
neutrinos. In particular, in the standard seesaw model, successful
leptogenesis implies a lower bound on the values of $M_{N_1}$ that, even in
the most favorable case of dominant $N_1$ initial abundance \cite{gi04}, puts
a direct test of leptogenesis out of the reach of any foreseeable
experiment.\footnote{Only in the case of resonant leptogenesis
  \cite{pi04-pi05}, scales much lower than the limit $M_{N_1}\gsim 10^7\,$GeV
  \cite{gi04} seem to be possible.}

In this paper we explore the implications of the presence below $M_{N_1}$, of
an additional energy scale related to the breaking of a new symmetry (that
could be for example a flavor symmetry).  As we will see, in some cases a
large enhancement of the $CP$ asymmetry in $N_1$ decays is easily obtained,
while at the same time the scale of leptogenesis can be lowered by several
orders of magnitude without conflicting with other conditions.  Indeed, the
scenario we will study can realize in a natural way some of the conditions
needed to render leptogenesis viable down to the TeV
scale~\cite{Hambye:2001eu}.  The main features of the model, that is directly
inspired by $U(1)$ models for flavor \`a la
Froggatt-Nielsen~\cite{Froggatt:1978nt} are outlined in section
\ref{sec:model}.\footnote{Leptogenesis models based on the Froggatt-Nielsen
  mechanism have been studied also in \cite{Cerdeno:2006ha} in the context of
  Dirac leptogenesis, and in \cite{Hambye:2004jf} as well as in the last paper
  in \cite{pi04-pi05} in the context of resonant leptogenesis.}  In section
\ref{sec:cp-asymmetries}, after reviewing the main results of standard
leptogenesis, we highlight the most important new features that stem from the
presence of the extra mass scale.  In section~\ref{sec:conclusions} we present
the conclusions. In the appendix we consider a minimal scenario where one
light neutrino remains massless, and we derive a simple analytical expression
for the ratio between the new mass scale and $M_{N_1}$.

\section{The model}
\label{sec:model}

We assume that at some large scale close to the leptogenesis scale a
horizontal $U(1)_X$ symmetry forbids direct couplings between the lepton
doublets $l$ and the heavy Majorana neutrinos $N$.  Light neutrino masses can
arise because of the presence of heavy vectorlike fields $F_L,\,F_R$ that are
singlets with respect to $SU(2)_L\times U(1)_Y$ and charged under the
additional $U(1)_X$ factor, and that couple to both the Majorana fields 
and to the lepton doublets $l$.  We use Greek indices $\alpha,\beta=1,\,2
\dots$ to denote the heavy Majorana neutrinos and we write the Lagrangian in
terms of self-conjugate Majorana spinors $N_\alpha \equiv ({N_R}_\alpha,
{N_R^c}_\alpha)^T$. Latin indices $a,\,b=1,\,2\dots $ will denote the heavy
Dirac  fields with $F_a \equiv ({F_R}_a,  {F_L}_a)^T$, and  Latin
indices $i,\,j=1,\,2,\,3$ will denote the SM left-handed lepton doublets
$l_i$.  The following Lagrangian gives a simple realization of this scheme:
\begin{eqnarray}
  \label{eq:lag}
  -{\cal L}&=& \frac{1}{2}\bar{N_\alpha} M_{N_{\alpha}} N_\alpha
  +\bar F_a M_{F_{a}} F_a +  
  \left(
    h_{ia}\overline{l}_{i}P_{R}F_{a}\Phi +
    \lambda_{\alpha a}\bar{N_{\alpha}} F_a S
  \right.
  \nonumber\\
  &&\hspace{6cm} \left.
    +\;\lambda^{(5)}_{\alpha a}\bar {N_{\alpha}} \gamma_5  F_a S+
    \mbox{h.c.}
  \right), 
\end{eqnarray}
where $\Phi$ is the $SU(2)_L$ Higgs doublet and $S$ is a complex scalar that
is a singlet under $SU(2)_L\times U(1)_Y$ and charged under $U(1)_X$. The
scalar $S$ is responsible for breaking the additional symmetry through a
vacuum expectation value (vev) $\sigma \equiv \langle S\rangle$ that we assume
to be somewhat smaller than the mass scale of the vectorlike fields $\sigma
\lsim M_F$.  A simple $U(1)_X$ charge assignment that forbids the $\bar l P_R N
\Phi$ coupling and yields the Lagrangian (\ref{eq:lag}) is for example
$X({l_L}_i,{F_L}_a,{F_R}_a)=+1$, $X(S)=-1$ and $X(N_\alpha,\Phi)=0$.

To put in evidence the main consequences of our scheme without complicating too
much the discussion and the results, we assume that to a good approximation
the heavy Dirac fields $F_a$ couple to the Majorana neutrinos in a pure
vectorlike way.  That is, we assume $\lambda^{(5)} \ll \lambda$, and  we
neglect all the effects related to the pseudoscalar couplings.  Including also
the pseudoscalar interactions would give rise to additional diagrams
contributing for example to the total decay width of the Majorana neutrinos
and to the decay $CP$ asymmetries, without changing our main conclusions.

Let us note that all the interaction terms in eq.~(\ref{eq:lag}) also preserve
a $U(1)$ (accidental) global symmetry with assignment
$L({l_L},{F_L},{F_R},N_R)=+1$ and $L(S,\Phi)=0$, that we can readily identify
with lepton number.  Then, as in the standard seesaw model, this symmetry is
broken (by two units) only by the $N_\alpha$ Majorana mass term.  As regards
the $U(1)_X$ symmetry, if for all the SM fields the charges $X$ were
proportional to $B-L$, this symmetry would leave unaffected all the charged
leptons and quarks Yukawa couplings. In contrast, if the set of $X$ charges
for the quarks and leptons is not a trivial one, $U(1)_X$ could play a role as
(part of) a flavor symmetry of the kind proposed long ago by Froggatt and
Nielsen~\cite{Froggatt:1978nt}.  However, for the present discussion the issue
if $U(1)_X$ also contributes to determine the charged fermion mass pattern is
to a large extent irrelevant, and accordingly we will not necessarily adopt
all the assumptions that realize the naturalness conditions of flavor models
based on Abelian symmetries, and in particular:
\begin{itemize} 
\item We do not constrain the couplings in the Lagrangian eq.~(\ref{eq:lag})
  to be all of the same size and of order unity.
\item We do not need to specify any precise value for the ratio 
  $\sigma/M_F$. 
\item We do not constrain from below by means of  FCNC considerations
the scale of $U(1)_X$ breaking (this is consistent if e.g.  
we take  $X=B-L$).
\end{itemize}

After $U(1)_X$ and electroweak symmetry breaking, the light neutrino mass
matrix arising from (\ref{eq:lag}) reads
\begin{equation}
  \label{eq:numatrix}
- {\cal M}_{ij} =  \left[h^*\frac{\sigma}{M_F}\lambda^T\frac{v^2}{M_N}
\lambda \frac{\sigma}{M_F} h^\dagger \right]_{ij} = 
\left[\tilde\lambda^T \frac{v^2}{M_N}\tilde\lambda\right]_{ij}
\end{equation}
where for convenience we have introduced the effective seesaw couplings 
\begin{equation}
  \label{eq:tildelambda}
\tilde\lambda_{\alpha i} = \left( \lambda
\frac{\sigma }{M_F}\,h^\dagger\right)_{\alpha i}.  
\end{equation}
That is, with respect to the standard seesaw mechanism the light neutrino
masses have an additional suppression factor of the ratio $\sigma^2/M_F^2$ and
are of fourth order in the fundamental couplings ($\lambda$ and $h$).

An important point to note is that in order to ensure that two light neutrino
are massive, as is required by  oscillation neutrino data, a minimum field
content of two right-handed neutrinos $N_{\alpha}$ and two vectorlike fields
$F_a$ is needed. A straightforward analysis then shows that even
in this minimal scheme, both the matrices of the $h$ and $\lambda$ coupling
constants contain physical complex phases that can be relevant for
leptogenesis.

To summarize,
in this model besides the electroweak breaking scale $v$ we have the following
new mass scales:

\begin{itemize}
\item the mass scale of the heavy vectorlike fields, $M_F$.
\item the lepton number breaking scale, $M_N$.
\item the horizontal symmetry breaking scale, $\sigma$.
\end{itemize}
The consequences of the different hierarchies amongst these new
scales is studied in detail in section \ref{sec:cp-asymmetries}.

\section{The  different possibilities}
\label{sec:cp-asymmetries}

In this section we present the detailed results for the $CP$ asymmetry in
$N_1$ decays for the different cases as are determined by the hierarchy
between the relevant scales $M_{N_1}$, $M_F$ and $\sigma$, and we will 
explore qualitatively the implications of the different possibilities. 
As already said, we take $N_1$ to be the lightest one of the Majorana
neutrinos, and $F_1$ to be the lightest of the vectorlike fields. For the
different mass ratios we adopt the following notation:
\begin{equation}
 z_{\alpha}=\frac{M_{N_\alpha}^2}{M_{N_1}^2},\qquad\qquad \omega_{a}=
\frac{M_{F_a}^2}{M_{F_1}^2}, \qquad\qquad r_a=\frac{M_{N_1}}{M_{F_a}}. 
\end{equation}
The $CP$-asymmetry in the decay of the heavy Majorana neutrinos $N_1$  
is defined in the usual way as
\begin{equation}
  \label{eq:cpdecay}
  \epsilon_{N_1}=\frac{\Gamma_{N_1}-\bar\Gamma_{N_1}}
{\Gamma_{N_1}+\bar\Gamma_{N_1}},
\end{equation}
where $\Gamma_{N_1}$ and $\bar \Gamma_{N_1}$ represent respectively the
partial decay rates of $N_1$ into particles with lepton number $L=+1$ and
antiparticles with lepton number $L=-1$  (regardless of the fact that they are
$l$ or $F$ states).  A non-vanishing numerator in eq.(\ref{eq:cpdecay}) can
arise from the interference between tree-level and loop-amplitudes, and the
total decay rate in the denominator can be approximated with the tree level
result.  When the ratio in eq.(\ref{eq:cpdecay}) is phase-space independent, as is
the case in two-body decays, the $CP$-asymmetry can be simply calculated 
in terms of products of the tree-level  ${\cal M}_0$  and loop ${\cal M}_1$ 
amplitudes. 
However, when there is a dependence on the phase space as is the case for
three-body decays discussed in section~\ref{sec:case1}, the full decay widths 
have to be taken into account.

In standard leptogenesis models the lepton number breaking scale is
constrained by the out-of-equilibrium condition on the decay rate of $N_1$, 
and as the scale of lepton number violation is lowered, in order to satisfy
this condition the size of the $N_1$ Yukawa couplings must be accordingly
reduced. If we further require that the neutrino oscillation data are
accounted for just by the (type 1) seesaw mass matrix, and we forbid any
additional source for the light neutrino masses, then it can be shown that a
large suppression of the $CP$ asymmetry in $N_1$ decays is unavoidable.  Then
the requirement that the final lepton asymmetry is large enough to account for
the baryon asymmetry of the Universe, implies a lower bound on the mass of
$N_1$ that is several orders of magnitude larger than the electroweak breaking
scale.

As we will discuss below, in some cases the presence of a new scale and of a
new set of couplings associated with it can allow to satisfy the
out-of-equilibrium condition and to account for the light neutrino mass scale,
without necessarily implying any particular suppression of the $CP$
asymmetries, even when $M_{N_1}$ is lowered down to the TeV scale.  This
`decoupling' of the size of the $CP$ asymmetry from the decay width
$\Gamma_{N_1}$ and from the scale of the light neutrino masses is rendered
possible by the fact that, while the latter two quantities are mainly
controlled by the $\lambda$ parameters that couple the heavy vectorlike fields
$F$ to the right handed neutrino $N_1$, the $CP$ asymmetry is essentially
determined by the couplings $h$ between the fermions $F$ and the lepton
doublets $l$.

\subsection{The standard leptogenesis case: $M_F,\,\sigma \gg M_{N}$ }
\label{sec:standard}

\begin{figure}[t]
\begin{center}
  \begin{pspicture}(-4,0.4)(6,3)
    \psline(-3.5,2)(-2.5,2)
    \psline[linestyle=dashed,dash=2pt 1.5pt]{->}(-2.5,2)(-2.2,2.3)
    \psline[linestyle=dashed,dash=2pt 1.5pt](-2.2,2.3)(-2.0,2.5)
    \psline{->}(-2.5,2)(-2.2,1.7)
    \psline(-2.2,1.7)(-2,1.5)
    \uput[d](-3.5,2){\footnotesize{$N_1$}}
    \uput[u](-2,2.5){\footnotesize{$\Phi$}}
    \uput[d](-2,1.5){\footnotesize{$l_j$}}
    \uput[d](-2.7,1){\footnotesize{$(a)$}}
      \psline(-1,2)(0,2)
      \psline[linestyle=dashed,dash=2pt 1.5pt]{->}(0,2)(0.3,2.3)
      \psline[linestyle=dashed,dash=2pt 1.5pt](0.25,2.25)(0.5,2.5)
      \psline(0,2)(0.3,1.7)
      \psline{<-}(0.15,1.85)(0.5,1.5)
      \psline(0.5,2.5)(0.5,1.5)
      \psline[linestyle=dashed,dash=2pt 1.5pt]{->}(0.5,1.5)(0.9,1.5)
      \psline[linestyle=dashed,dash=2pt 1.5pt](0.9,1.5)(1.2,1.5)
      \psline{->}(0.5,2.5)(1,2.5)
      \psline(0.9,2.5)(1.2,2.5)
      \uput[d](-0.8,2){\footnotesize{$N_1$}}
      \uput[l](0.3,2.4){\footnotesize{$l_i$}}
      \uput[l](0.3,1.5){\footnotesize{$\Phi$}}
      \uput[r](0.4,2){\footnotesize{$N_\beta$}}
      \uput[d](1.1,1.5){\footnotesize{$\Phi$}}
      \uput[u](1.1,2.4){\footnotesize{$l_j$}}
      \uput[d](0.5,1){\footnotesize{$(b)$}}
      \psline(2.1,2)(2.8,2)
      \psarc{->}(3.2,2){0.4cm}{0}{90}
      \psarc(3.2,2){0.4cm}{90}{180}
      \psarc[linestyle=dashed,dash=2pt 1.5pt]{->}(3.2,2){0.4cm}{180}{270}
      \psarc[linestyle=dashed,dash=2pt 1.5pt](3.2,2){0.4cm}{270}{360}
      \psline(3.6,2)(4.1,2)
      \psline{->}(4.1,2)(4.4,2.3)
      \psline(4.3,2.2)(4.6,2.5)
      \psline[linestyle=dashed,dash=2pt 1.5pt]{->}(4.1,2)(4.4,1.7)
      \psline[linestyle=dashed,dash=2pt 1.5pt](4.4,1.7)(4.6,1.5)
      \uput[u](3.2,2.4){\footnotesize{$l_i$}}
      \uput[d](3.2,1.6){\footnotesize{$\Phi$}}
      \uput[d](2.3,2){\footnotesize{$N_1$}}
      \uput[d](3.9,2){\footnotesize{$N_\beta$}}
      \uput[d](4.6,1.5){\footnotesize{$\Phi$}}
      \uput[u](4.6,2.4){\footnotesize{$l_j$}}
      \uput[d](3.2,1){\footnotesize{$(c)$}}
  \end{pspicture}
\end{center}
\caption{Diagrams generating the lepton asymmetry in the
  Fukugita-Yanagida model.}
\label{fig:case0}
\end{figure}
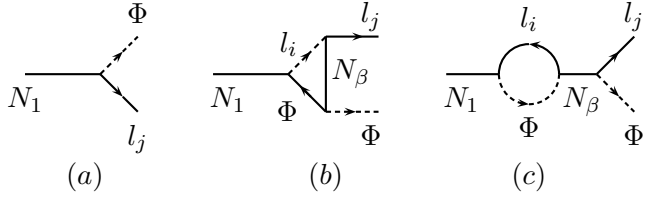

When the masses of the heavy Dirac fields and the $U(1)_X$ symmetry breaking
scale are both larger than the Majorana neutrino masses ($M_F\,, \sigma >
M_N$) there are no major differences from the standard Fukugita-Yanagida
leptogenesis model \cite{fu86}.  After integrating out the $F$ fields one
obtains the standard seesaw Lagrangian containing the effective operators
$\tilde\lambda_{\alpha i} \bar N_\alpha l_i \Phi$ with the seesaw couplings
$\tilde\lambda_{\alpha i}$ given in eq.~(\ref{eq:tildelambda}).  The right
handed neutrino $N_1$ decays predominantly via 2-body channels as shown in
fig.~\ref{fig:case0}.  This yields the standard results that for convenience
we recall here.  The total decay width is $\Gamma_{N_1}= \left({M_{N_1}}/{16
    \pi}\right) (\tilde\lambda\tilde\lambda^\dagger)_{11}$ and the sum of the
vertex and self-energy contributions to the $CP$-asymmetry for $N_1$ decays
into the flavor $l_j$ reads \cite{co96}
\begin{equation}
        \label{eq:6} 
\epsilon_{N_1\to l_j}=\frac{1}{8\pi (\tilde\lambda \tilde\lambda^\dagger)_{11}}
\sum_{\beta\neq 1}{\mathbb I}\mbox{m}
\left\{\tilde\lambda_{\beta j}\tilde\lambda^*_{1j} \left[
(\tilde\lambda\tilde\lambda^\dagger)_{\beta1}    \tilde F_1(z_\beta) +
(\tilde\lambda\tilde\lambda^\dagger
       )_{1\beta}
    \tilde F_2(z_\beta)
\right]\right\}, 
\end{equation}
where 
%
\begin{equation}
  \label{eq:tildeF}
      \tilde F_1(z )= \frac{\sqrt{z }}{1-z } +
  \sqrt{z }\left(1-(1+z )\ln
    \frac{1+z }{z }\right),   
\qquad     \tilde F_2(z)=\frac{1}{1-z}. 
\end{equation}
%
At leading order in $1/z_\beta$ and after summing over all leptons $l_j$,    
eq.~(\ref{eq:6}) yields for the total asymmetry: 
\begin{equation}
        \label{eq:8} 
\epsilon_{N_1}=\frac{3}{16\pi (\tilde\lambda \tilde\lambda^\dagger)_{11}}
\sum_{\beta}
{\mathbb I}\mbox{m}
\left\{\frac{1}{\sqrt{z_\beta}}
       (\tilde\lambda\tilde\lambda^\dagger)_{\beta1}^2 \right\}. 
\end{equation}
where the sum over the heavy neutrinos has been extended to include also $N_1$
since  for $\beta=1$  the corresponding combination of couplings is real.

In the hierarchical case $M_{N_1}\ll M_{N_{2,3}}$  
the size of the total asymmetry in (\ref{eq:8}) is bounded by 
the Davidson-Ibarra limit  \cite{Davidson:2002qv}
\begin{equation}
\label{eq:DI}
|\epsilon_{N_1}|\leq \frac{3}{16\pi}\frac{M_{N_1}}{v^2} \,(m_{\nu_3}-m_{\nu_1})
\lsim \frac{3}{16\pi}\frac{M_{N_1}}{v^2} \frac{\Delta m^2_{atm}}{2m_{\nu_3}}\,, 
\end{equation}
where $m_{\nu_i}$ (with $m_{\nu_1} < m_{\nu_2} < m_{\nu_3}$) are the light
neutrinos mass eigenstates and $\Delta m^2_{atm}\sim 2.5\times
10^{-3}\,$eV$^2$ is the atmospheric neutrino mass difference.  It is now easy
to see that (\ref{eq:DI}) implies a lower limit on $M_{N_1}$.  The amount of
$B$ asymmetry that can be generated from $N_1$ dynamics can be written as:
\begin{equation}
\label{eq:etaB}
\frac{n_B}{s}=-\kappa_s\,\epsilon_{N_1}\,\eta  ,
\end{equation} 
where $\kappa_s\approx 1.3\times 10^{-3}$ accounts for the dilution of the
asymmetry due to the increase of the Universe entropy from the time the
asymmetry is generated with respect to the present time, $\eta $ (that can
range between 0 and 1, with typical values $10^{-1}-10^{-2}$) is the {\it
efficiency factor} that accounts for the amount of $L$ asymmetry that can
survive the washout process.  WMAP data on the cosmic background anisotropy
\cite{Spergel:2006hy} and considerations of big bang nucleosynthesis
\cite{Steigman:2005uz} yield the experimental value ${n_B}/{s} \approx (8.7\pm
.4)\times 10^{-11}$, and therefore, assuming that $\epsilon_{N_1}$ 
is the main source of the $B-L$ asymmetry~\cite{eng06}, 
 eqs.~(\ref{eq:DI}) and (\ref{eq:etaB})  yield: 
\begin{equation}
  \label{eq:M1limit0}
  M_{N_1} \gsim 10^{9}\, 
\frac{m_{\nu_3}}{\eta\, \sqrt{\Delta m^2_{atm}}} \, {\rm GeV}.
\end{equation}
This limit can be somewhat relaxed depending on the specific initial
conditions~\cite{gi04} or when flavor effects are
included~\cite{aba06a-aba06b,Josse-Michaux:2007zj} but the main point remains,
and that is that the value of $M_{N_1}$ should be well above the electroweak
scale.

%

\subsection{Variations on leptogenesis: $\sigma < M_{N_1}$ }

Interesting new possibilities arise when the $U(1)_X$ symmetry breaking scale
is lower than the leptogenesis scale, that is $\sigma < M_{N_1}$. In this case
$N_1$ cannot decay directly into the light lepton doublets via the two body
channel.  In this regime, we can distinguish three cases:

\begin{itemize} \itemsep 4pt
  
\item[1)] $\sigma < M_{N_1}$ but all the masses $M_{F_a}$ are larger than
  $M_{N_1}$: then $N_1$ can decay to $l_j$ only via the three body channel
  $N_1\to S\, \Phi\, l_j$  depicted in figure~\ref{fig:case1}.
  
\item[2)] $M_{N_1}$ is larger than all the scales related to the $U(1)_X$
  symmetry ($M_F,\,\sigma$): then $N_1$ will decay via two body channels to
  $F_a$ and $\bar F_a$, (see figure~\ref{fig:case2}). The heavy fermions $F$
  will then transfer part of the asymmetry to the light leptons via lepton
  number conserving processes (decays and scatterings).
  
\item[3)] If $M_N$ arises from the same source than $M_F$ (as for example from
  the vev of a singlet) then some of the heavy fermions (for example $F_1$)
  could be lighter than $N_1$ while the others can be heavier.  Then $N_1$
  will decay dominantly into $F_1,\bar F_1$ via the two body channel in
  figure~\ref{fig:case2}. However, a new diagram contributing to the $CP$
  asymmetry is present in this case.  This diagram is interesting, since it
  yields the possibility of decoupling the lifetime of $N_1$ from the size of
  the $CP$ violating asymmetry $\epsilon_{N_1\to F_1}$.

\end{itemize}

In the next sections we will analyze in some detail these different 
possibilities.

\subsubsection{Case 1: $M_F > M_{N_1} $ }
\label{sec:case1}

\begin{figure}[h]
  \centering
  \begin{pspicture}(-4,-1.5)(4,2)
    \psline{->}(-4.5,0)(-3,0)
    \uput[d](-4.3,0){\footnotesize{$N_1$}}
    \psline{->}(-3,0)(-2,0)
    \uput[d](-3,0){\footnotesize{$F_a$}}
    \psline(-2,0)(-1.5,0)
    \uput[d](-1.5,0){\footnotesize{$l_j$}}
    \psline[linestyle=dashed,dash=2pt 1.5pt](-3.7,1)(-3.7,0.6)
    \psline[linestyle=dashed,dash=2pt 1.5pt](-3.7,0.6)(-3.7,0)
    \uput[u](-3.7,1){\footnotesize{$S$}}
    \psline[linestyle=dashed,dash=2pt 1.5pt](-2.5,1)(-2.5,0.6)
    \psline[linestyle=dashed,dash=2pt 1.5pt](-2.5,0.6)(-2.5,0)
    \uput[u](-2.5,1){\footnotesize{$\Phi$}}
    \uput[d](-3,-1){\footnotesize{$(a)$}}
    \psline{->}(0,0)(1.3,0)
      \psline(1.2,0)(1.7,0)
      \psline[linestyle=dashed,dash=2pt 1.5pt](0.7,0)(0.7,0.6)
      \psline[linestyle=dashed,dash=2pt 1.5pt](0.7,0.6)(0.7,1)
      \psarc(2.1,0){0.4cm}{0}{90}
      \psarc{<-}(2.1,0){0.4cm}{90}{180}
      \psarc[linestyle=dashed,dash=2pt 1.5pt](2.1,0){0.4cm}{180}{270}
      \psarc[linestyle=dashed,dash=2pt 1.5pt](2.1,0){0.4cm}{270}{360}
      \psline{->}(2.5,0)(3.2,0)
      \psline(3.1,0)(3.6,0)
      \psline[linestyle=dashed,dash=2pt 1.5pt](3.5,0)(3.5,0.6)
      \psline[linestyle=dashed,dash=2pt 1.5pt](3.5,0.6)(3.5,1)
      \psline{->}(3.5,0)(3.9,0)
      \psline(3.8,0)(4.2,0)
      \uput[d](0.2,0){\footnotesize{$N_1$}}
      \uput[d](1.1,0){\footnotesize{$F_{a'}$}}
      \uput[u](0.8,1){\footnotesize{$S$}}
      \uput[u](2.1,0.4){\footnotesize{$l_m$}}
      \uput[d](2.1,-0.4){\footnotesize{$\Phi$}}
      \uput[d](3.1,0){\footnotesize{$F_{b}$}}
      \uput[d](4.2,0){\footnotesize{$l_j$}}
      \uput[u](3.5,1){\footnotesize{$\Phi$}}
      \uput[d](2.1,-1){\footnotesize{$(b)$}}
    \end{pspicture}
  \caption{Diagrams responsible for the $CP$-asymmetry in case 1.}
  \label{fig:case1}
\end{figure}
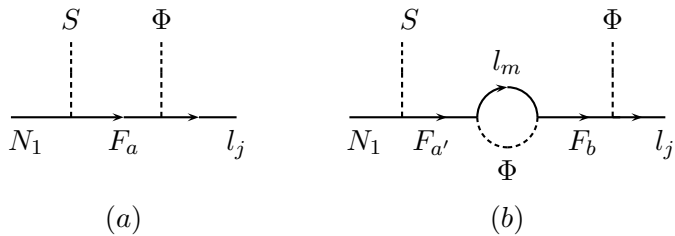

In the case when all the vectorlike fermions $F$ are heavier than $N_1$ but
the horizontal $U(1)_X$ is still a good symmetry at $T\lsim M_{N_1}$, there is
only one diagram contributing to the $CP$ asymmetry, that is the wave function
type of diagram depicted in figure~\ref{fig:case1}(b).  This is because on the
one hand the loop correction to the $N_1\,F\,S$ vertex with $N_\beta$
($\beta\neq 1$) and $F$ internal lines does not develop an imaginary part, and
on the one other hand there is no correction at one loop to the vertex
$F\,l\,\Phi$ (the first correction arises only at two-loops).  This is due to
the fact that the standard vertex correction depicted in
figure~\ref{fig:case0}$(b)$ is of the Majorana type (with an inverted flow of
fermion number) while in the present case both $F$ and $l$ are Dirac fermions.

Unlike the standard case and the next two cases discussed in
sections~\ref{sec:case2} and \ref{sec:case3}, for three body final states the
squared amplitudes and interference terms are not phase-space independent. Our
results for the total decay width and $CP$-asymmetries given below correspond
to the leading terms in the mass ratios $r_{a,a',b} < 1$.  For the $N_1$ total
decay width we obtain
\begin{equation}
  \label{eq:width-decay-treel}
  \Gamma_{N_1}=\frac{M_{N_1}}{192\,\pi^3}
  \sum_{a\,a'}\,r_{a'}\,r_a\,
  \left(h^\dagger h\right)_{a'a}\,\lambda_{1a'}\lambda_{1a}^*
= \frac{M_{N_1}}{192\,\pi^3} \left(\frac{M_{N_1}}{\sigma  }\right)^2 
(\tilde\lambda\tilde\lambda^\dagger)_{11}\,.
\end{equation}
The interference between diagrams \ref{fig:case1}$(a)$ and
\ref{fig:case1}$(b)$, summed over all the leptons $l_m$ running in the loop,
yields the $CP$-asymmetry:
\begin{eqnarray}
  \label{eq:cpasymmetrythreebody}
\nonumber
  \epsilon_{N_1}&=&
  \frac{3}{128\pi}
  \frac{  \sum_{a\,a'b}\>
{\mathbb I}\mbox{m}
\left[
(h^\dagger h)_{a'b}\,(h^\dagger h)_{ba}\,
\lambda_{1a'}\,\lambda_{1a}^*
\right]\, r_{a'}\,r_a\,r_b^2\, } 
  {\sum_{a\,a'}\>\left(h^\dagger h\right)_{a'a}\left(\lambda_{1a'}
    \lambda_{1a}^*\right)\,r_a\,r_{a'} }\\
&=&   \frac{3}{128\pi}   \frac{  
{\mathbb I}\mbox{m} [\,\tilde\lambda\, h\, r^2\,
  h^\dagger\,\tilde\lambda^\dagger]_{11} }
{(\tilde \lambda\tilde \lambda^\dagger)_{11}}
= 0.
\end{eqnarray}
When written in terms of the effective couplings $\tilde\lambda$ as in the
second line in eq.~(\ref{eq:cpasymmetrythreebody}), the vanishing of the $CP$
asymmetry at leading order in the mass ratios $r_{a,a',b} < 1$ is apparent.
However, this is true also beyond the leading order, and is related to the
fact that the loop in diagram~\ref{fig:case1}$(b)$ does not involve lepton
number violation.  Mathematically, one can see from the first equality in
eq.~(\ref{eq:cpasymmetrythreebody}) that the vanishing of the $CP$-asymmetry
follows from the fact that by exchanging the two indices $a \leftrightarrow
a'$ the combination of couplings within square brackets goes into its complex
conjugate, and thus the sum over $a$ and $a'$ yields a real term for any value
of the index $b$.  Since the full phase-space function (that in
eq.~(\ref{eq:cpasymmetrythreebody}) has been replaced with the leading term
$\sim r_a r_a'$) is symmetric under the exchange $ r_a \leftrightarrow r_a'$
this holds also for the complete result.  The fact that there is no source
term for the lepton asymmetry implies that in the present case leptogenesis
can occur only through the effects of lepton flavor dynamics combined with the
violation of lepton number that is provided by the washout processes.  The
single lepton-flavor $CP$-asymmetries are indeed non-vanishing:
\begin{eqnarray}
  \label{eq:cpasymmetrythreebodyflavor}
\nonumber
  \epsilon_{N_1\to l_j}&=&
 \frac{3}{128\pi}
  \frac{  \sum_{a\,a'\,b}\>
{\mathbb I}\mbox{m}
\left[
(h^\dagger h)_{a'b}
\,h^*_{jb} h_{ja}\,
\lambda_{1a'}\,\lambda_{1a}^*
\right]\,r_{a'}\,r_a\,r_b^2\, } 
  {\sum_{a\,a'}\;\left(h^\dagger h\right)_{a'a}\lambda_{1a'}
    \lambda_{1a}^*\,r_{a'}\,r_a} \\
&=&
 \frac{3}{128\pi}   \frac{\sum_i  
{\mathbb I}\mbox{m} \left[\left(h r^2
  h^\dagger\right)_{ij} \tilde\lambda_{1i}\tilde\lambda^*_{1j}\right]}
{\left(\tilde \lambda\tilde \lambda^\dagger\right)_{11}}
\neq 0\,.
\end{eqnarray}
and will be a source of non vanishing asymmetry-densities in the different
flavors. These asymmetries will then suffer washouts processes (like $s$ and
$t$ channels scatterings $ l_j\Phi \leftrightarrow S^{*} N_1$, $S N_1
\leftrightarrow \bar l_j \bar \Phi$) that in general are characterized by
different rates for the different flavors.  As is discussed e.g. in
ref.~\cite{na06}, under these conditions a net lepton asymmetry can result
even if $\epsilon_{N_1}=0$.

A quick inspection of equation (\ref{eq:cpasymmetrythreebodyflavor}) shows
that if the $\tilde \lambda$ couplings are all of the same order of magnitude,
the $CP$ asymmetry $ \epsilon_{N_1\to l_j}$ is roughly proportional to $ |h r^2
h^{\dagger}|$.  That is, the dominant contribution to the $CP$ asymmetry is
determined by the $h$ couplings that are different from the effective
couplings $\tilde \lambda$ appearing in the neutrino mass matrix, and in
particular larger by a factor of $(\lambda \sigma/M_F)^{-1}$ (see
eq.~(\ref{eq:tildelambda})).  If we further assume the hierarchy $h r > \lambda$ between the fundamental
couplings then, independently of the particular
value of $M_{N_1}$, leptogenesis will always occur in a regime when the $F_a$
interactions with the light leptons determine a `flavor' basis $\ell_a =
h_{ia} l_i/\sqrt{(hh^\dagger)_{aa}}$.  This ensures that the requirement that
flavor dynamics participates in the generation of a lepton asymmetry is satisfied.

We will now address the following two interesting points: 1) In the present case is leptogenesis still compatible with a reasonable scale for the light neutrino masses?  2) Is the leptogenesis scale still bounded from below, or
can it be lowered to values that are experimentally accessible?

The out-of-equilibrium condition necessary to ensure that a macroscopic lepton
asymmetry can be generated reads 
\begin{equation}
  \label{eq:outofeq-general}
  \Gamma_{N_1}\lsim \xi\cdot H(M_{N_1}), 
\end{equation}
where the out-of-equilibrium parameter $\xi$ can normally lie in the range
$\xi \sim 0.1-10$ (but values as large as $10^{2}$ are possible) and the
Hubble parameter at decay time is $H(M_{N_1})\simeq 1.66 \sqrt{g_*}
M^2_{N_1}/M_P$ with $g_*=106.75$ the number of relativistic degrees of freedom
at $T\sim M_{N_1}$ and $M_P$ the Planck mass.  Using
(\ref{eq:width-decay-treel}) this yields the condition
\begin{equation}
  \label{eq:out-of-eq}
  (\tilde\lambda \tilde\lambda^\dagger)_{11}\lsim 10^5\> \xi  
  \left(\frac{\sigma}{M_{N_1}}\right)^2
\frac{M_{N_1}}{M_P}. 
\end{equation}
{}From equation (\ref{eq:numatrix}) we have 
\begin{equation}
  \label{eq:trace}
\sum_i m_{\nu_i}= \frac{v^2}{M_{N_1}}  {\rm Tr} 
\left(\tilde\lambda^T z^{-1/2}\tilde\lambda\right)_{11}
\approx \frac{v^2}{M_{N_1}} (\tilde\lambda\tilde\lambda^T)_{11},
\end{equation}
where in the second relation we have assumed that $M_{N_1}$ dominates the
seesaw matrix. From (\ref{eq:out-of-eq}) and (\ref{eq:trace}) we obtain the
order-of-magnitude relation
\begin{equation}
  \label{eq:numassess}
  \sum_i m_{\nu_i} \approx 0.3 \> \xi \,
  \left(\frac{\sigma}{M_{N_1}}\right)^2\,{\rm eV}.
\end{equation}
This ensures that if the ratio $\sigma/M_{N_1}$ is not exceedingly small, the
out-of-equilibrium condition can be satisfied for the correct scale of
neutrino masses.  When the value $\xi \cdot \left(\sigma/M_{N_1}\right)^2 \sim
10^{-1}$ suggested by the previous equation is inserted into
eq.~(\ref{eq:out-of-eq}) we obtain $ (\tilde\lambda
\tilde\lambda^\dagger)_{11}\lsim 10^{-12} \,(M_{N_1}/ 1\,{\rm TeV})$, that is
the $\tilde\lambda$ couplings should be of the order of the electron Yukawa
coupling when $M_{N_1}$ is at the TeV scale.  On the other hand this limit
does not constrain the size of the asymmetry in
eq.~(\ref{eq:cpasymmetrythreebodyflavor}), and for $|hr^2h^\dagger|\gsim
10^{-3}$ the $CP$ asymmetry could be sufficiently large for successful
leptogenesis.  For such a low leptogenesis scale, direct production of the $F$
states could be possible in collider experiments e.g. via off-shell $X$ boson
exchange ($M_X\sim g_X \sigma < M_F$). However, $F\to N_1$ decays (if
kinematically accessible) would be strongly suppressed by the small values of
$\lambda$, and thus a direct detection of the Majorana neutrinos would be a
rather difficult task.

\subsubsection{Case 2: $ M_F < M_{N_1} $ }
\label{sec:case2}

\begin{figure}[t!]
  \centering
  \begin{pspicture}(-3,0.4)(4,3)
    \psline(-3.5,2)(-2.5,2)
    \psline[linestyle=dashed,dash=2pt 1.5pt](-2.5,2)(-2.2,1.7)
    \psline[linestyle=dashed,dash=2pt 1.5pt](-2.2,1.7)(-2,1.5)
    \psline{->}(-2.5,2)(-2.2,2.3)
    \psline(-2.2,2.3)(-2.0,2.5)
    \uput[d](-3.5,2){\footnotesize{$N_1$}}
    \uput[u](-2,2.5){\footnotesize{$F_a$}}
    \uput[d](-2,1.5){\footnotesize{$S$}}
    \uput[d](-2.7,1){\footnotesize{$(a)$}}
      \psline(-1,2)(0,2)
      \psline[linestyle=dashed,dash=2pt 1.5pt](0,2)(0.3,2.3)
      \psline[linestyle=dashed,dash=2pt 1.5pt](0.3,2.3)(0.5,2.5)
      \psline{->}(0,2)(0.3,1.7)
      \psline(0.3,1.7)(0.5,1.5)
      \psline(0.5,2.5)(0.5,1.5)
      \psline[linestyle=dashed,dash=2pt 1.5pt](0.5,1.5)(0.9,1.5)
      \psline[linestyle=dashed,dash=2pt 1.5pt](0.9,1.5)(1.2,1.5)
      \psline{->}(0.5,2.5)(1,2.5)
      \psline(0.9,2.5)(1.2,2.5)
      \uput[d](-0.8,2){\footnotesize{$N_1$}}
      \uput[l](0.4,2.5){\footnotesize{$S$}}
      \uput[l](0.5,1.5){\footnotesize{$F_b$}}
      \uput[r](0.4,2){\footnotesize{$N_\beta$}}
      \uput[d](1.1,1.5){\footnotesize{$S$}}
      \uput[u](1.1,2.4){\footnotesize{$F_a$}}
      \uput[d](0.5,1){\footnotesize{$(b)$}}
      \psline(2.1,2)(2.8,2)
      \psarc(3.2,2){0.4cm}{0}{90}
      \psarc{<-}(3.2,2){0.4cm}{90}{180}
      \psarc[linestyle=dashed,dash=2pt 1.5pt](3.2,2){0.4cm}{180}{270}
      \psarc[linestyle=dashed,dash=2pt 1.5pt](3.2,2){0.4cm}{270}{360}
      \psline(3.6,2)(4.1,2)
      \psline{->}(4.1,2)(4.4,2.3)
      \psline(4.3,2.2)(4.6,2.5)
      \psline[linestyle=dashed,dash=2pt 1.5pt](4.1,2)(4.4,1.7)
      \psline[linestyle=dashed,dash=2pt 1.5pt](4.4,1.7)(4.6,1.5)
      \uput[u](3.2,2.4){\footnotesize{$F_b$}}
      \uput[d](3.2,1.6){\footnotesize{$S$}}
      \uput[d](2.3,2){\footnotesize{$N_1$}}
      \uput[d](3.9,2){\footnotesize{$N_\beta$}}
      \uput[d](4.6,1.5){\footnotesize{$S$}}
      \uput[u](4.6,2.4){\footnotesize{$F_a$}}
      \uput[d](3.2,1){\footnotesize{$(c)$}}
  \end{pspicture}
  \caption{Diagrams generating the $CP$ 
    asymmetry for $M_{N_1}> M_F,\, \sigma $ }
  \label{fig:case2}
\end{figure}
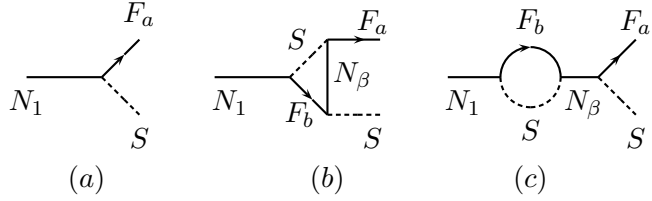

In this case $N_1$ decays proceed through the diagram in
figure~\ref{fig:case2}$(a)$.  A lepton asymmetry is first generated in the $F$
states, and is transferred in part to the light leptons through the $L$
conserving interactions controlled by the couplings $h$.  If we assume that
the $h$-interactions are in equilibrium at $T\sim M_{N_1}$ (as is the case if
the couplings $h$ are larger than the couplings $\lambda_{1a}$), then the
reduction of the asymmetry in the $F$ states implied by chemical equilibrium
with the light leptons $l$ also implies a reduction in the rates of the washout
processes, with a corresponding enhancement of the efficiency $\eta$.  This
would favor the survival of a sizeable asymmetry.  Neglecting in first
approximation phase space suppressions from final state masses, the total
decay width for this case reads:
\begin{equation}
  \label{eq:width1}
\Gamma_{N_1}=\frac{M_{N_1}}{32\pi} \left(\lambda\lambda^\dagger\right)_{11}.    
\end{equation}
The $CP$ asymmetry is determined by the interference 
between diagram $(a)$ and the loop diagrams $(b)$ and $(c)$  
in figure~\ref{fig:case2}.
%
Note that even if these diagrams are the same as in the standard case, the
result for the asymmetry is different, since in contrast to the lepton
doublets the vectorlike fields $F$ do not couple chirally to the Majorana
neutrinos.  From the interference of diagrams {\it (a)} and {\it (b)} in
figure~\ref{fig:case2} we obtain:
\begin{equation}
  \label{eq:case2-diagb-6}
  \epsilon^{(a+b)}_{N_1\to F_a}=
  \frac{1}{8\pi (\lambda\lambda^\dagger)_{11}}
  \sum_{\beta\neq 1}{\mathbb I}\mbox{m}
  \left[(\lambda\lambda^\dagger)_{\beta1}\,\lambda_{\beta 
      a}\lambda^*_{1 a}\right]
  F_1(z_\beta)
\end{equation}
where  $F_1$ given by
\begin{eqnarray}
  \label{eq:case2-diagb-7}
  \nonumber
  F_1(z )&=& -2z -\sqrt{z }
  +\left[z (1+2z )
  +\sqrt{z }(1+z )\right]
  \log\left(    \frac{1+z }{z }
  \right)\ \nonumber \\
& \stackrel{z  \to \infty}{=} &\ 
\frac{1}{2\,\sqrt{z }}+\frac{1}{6\,z }+\dots 
\end{eqnarray}
As regards the self-energy type of diagrams, like in the standard case besides
the diagram depicted in figure~\ref{fig:case2}$(c)$ there is another diagram
$(c')$ of the Majorana type with opposite fermion flux in the loop.  The
contribution from the interference of these diagrams with diagram $3(a)$ is:
\begin{equation}
  \label{eq:cas22-diagc-2}
  \epsilon^{(a+c+c')}_{N_1\to F_a}=
   \frac{-1}{8\pi(\lambda\lambda^\dagger)_{1 1}}\sum_{\beta\neq 1}  
  {\mathbb I}\mbox{m}\left\{
    \left[(\lambda\lambda^\dagger)_{\beta1}+
      (\lambda\lambda^\dagger)_{1\beta}
    \right]\lambda_{\beta a}\lambda^*_{1a}\right\}\, F_2(z_\beta)
\end{equation}
with 
 \begin{equation}
  \label{eq:cas22-diagc-F}
F_2(z)=\frac{1+\sqrt{z}}{1-z} \  \stackrel{z \to
  \infty}{=} \ 
-\frac{1}{\sqrt{z}}-\frac{1}{z}+\dots 
\end{equation}
Adding the two contributions (\ref{eq:case2-diagb-6}) and
(\ref{eq:cas22-diagc-2}), summing over all the possible final states $F_a$,
and using the fact that $ {\mathbb
  I}\mbox{m}\left[(\lambda\lambda^\dagger)_{1\beta}
  (\lambda\lambda^\dagger)_{\beta1}\right] =0$ to include in the leading
order result also the sum over $\beta=1$, we obtain:
\begin{equation}
  \label{eq:epsilon-case2}
  \epsilon_{N_1} =
   \frac{3}{16\pi(\lambda\lambda^\dagger)_{1 1}}\sum_\beta 
  {\mathbb I}\mbox{m} \left\{\frac{1}{\sqrt{z_\beta}}
(\lambda\lambda^\dagger)_{\beta 1}^2 \right\}.
\end{equation}
Even if the two functions $F_1$ and $F_2$ are not the same as in the standard
case with chiral leptons couplings \cite{co96}, once the effective couplings
$\tilde\lambda$ are replaced with the fundamental $\lambda$'s, at leading
order eq.~(\ref{eq:epsilon-case2}) coincides with the standard expression
eq.~(\ref{eq:8}).  However, the fact that the couplings in eq.~
(\ref{eq:epsilon-case2}) are not the seesaw couplings $\tilde \lambda$ is
rather important, because the fundamental $\lambda$'s do not directly determine
the light neutrinos masses. This implies that for this case the bound
of eq.~(\ref{eq:DI}) does not hold. As regards the consistency among different
phenomenological requirements, an order of magnitude estimate suggests that
there is some tension between the out-of-equilibrium condition
(\ref{eq:outofeq-general}), that by means of eq.~(\ref{eq:width1}) yields
$(\lambda\lambda^\dagger)_{11} \lsim 10^3\,\xi\, M_{N_1}/M_P$, and the light
neutrino mass scale. In fact the latter can get too much suppression from the
small values of $\tilde\lambda$ that are reduced by the factor $h \sigma/M_F$
with respect to the $\lambda$ that determine the decay rate.  This yields the
rough estimate $\sum_i m_{\nu_i} \sim {\cal O} (\xi\,10^{-5})\,$eV. While it
is always possible to assume a large hierarchy between the combination of
couplings that control the out-of-equilibrium condition and the couplings
that determine the light neutrino masses, the fair conclusion is that in the
case under discussion successful leptogenesis can be ensured only by means of
a careful choice of the relevant parameters.

\subsubsection{Case 3 ($ M_{F_{2,3}}> M_{N_1} > M_{F_1} $)} 
 \label{sec:case3}

\begin{figure}[t]
  \centering
  \begin{pspicture}(-4,2.3)(0,5)
      \psline{->}(-4,3.7)(-2.7,3.7)
      \psline(-2.8,3.7)(-2.3,3.7)
      \psline[linestyle=dashed,dash=2pt 1.5pt](-3.3,3.7)(-3.3,4.1) 
      \psline[linestyle=dashed,dash=2pt 1.5pt](-3.3,4.1)(-3.3,4.5) 
      \psarc(-1.9,3.7){0.4cm}{0}{90} 
      \psarc{<-}(-1.9,3.7){0.4cm}{90}{180} 
      \psarc[linestyle=dashed,dash=2pt 1.5pt](-1.9,3.7){0.4cm}{180}{270} 
      \psarc[linestyle=dashed,dash=2pt 1.5pt](-1.9,3.7){0.4cm}{270}{360} 
      \psline{->}(-1.5,3.7)(-.8,3.7)
      \psline(-.9,3.7)(-.4,3.7)
      \uput[d](-3.8,3.7){\footnotesize{$N_1$}}
      \uput[d](-2.7,3.7){\footnotesize{$F_b$}}
      \uput[u](-3.3,4.5){\footnotesize{$S$}}
      \uput[u](-1.9,4.1){\footnotesize{$l_i$}}
      \uput[d](-1.9,3.3){\footnotesize{$\Phi$}}
      \uput[d](-1.9,2.7){\footnotesize{$(d)$}}
      \uput[d](-.7,3.7){\footnotesize{$F_1$}}
  \end{pspicture}
  \caption{Additional diagram contributing to the $CP$ asymmetry 
in $N_1\to F_1$ decays when $M_{F_1} < M_{N_1} <M_{F_{b}}$ ($b=2,3,\dots$).}
  \label{fig:case3}
\end{figure}
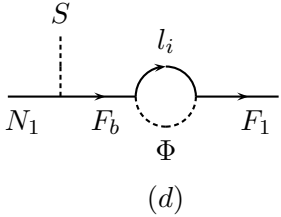

\noindent
This case corresponds to the situation when the value of $M_{N_1}$ lies in
between different values of $M_{F_a}$.  
For definiteness we assume $M_{F_1}< M_{N_1} < M_{F_b}$ with $b\neq 1$.
Neglecting the contributions from three body decays $N_1\to S\,l_i\,\Phi$
(suppressed by ${\cal O}(hh^*)$ and by phase space factors) 
the $N_1$ decay rate reads
\begin{equation}
  \label{eq:decaycase4}
\Gamma_{N_1}=\frac{M_{N_1}}{96\pi} |\lambda_{11}|^2.        
\end{equation}
For the $CP$ asymmetry, besides the diagrams in figure~\ref{fig:case2}$(a)$
the new type of diagram depicted in figure~\ref{fig:case3}$(d)$ also contributes.
This contribution is qualitatively different from the previous ones in
figure~\ref{fig:case2} since it involves the couplings $h_{ia}$ of the light
leptons $l_i$ to the vectorlike fermions $F_a$. It is easy to see that in the
case when the $h$ couplings dominate over the $\lambda$,
diagram~\ref{fig:case3}$(d)$ gives the leading contribution to the $CP$
asymmetry. The interference between diagrams ~\ref{fig:case2}$(a)$ and
~\ref{fig:case3}$(d)$ yields:
\begin{equation}
  \label{eq:cp-asymm-case3}
  \epsilon^{(a+d)}_{N_1\to F_1}=\frac{1}{8\pi |\lambda_{11}|^2}
  \sum_{b\neq 1}
{\mathbb I}\mbox{m}\left[   
(h^\dagger h)_{1b} \lambda_{11}\lambda_{1b}^{*} \right]
F_2(\omega_{b}) \,,
\end{equation}
where the function $F_2$ is given in eq.~(\ref{eq:cas22-diagc-F}). After
approximating $F_2(\omega)\sim -1/\sqrt{\omega}$) the sum can be extended over
all $b$ since ${\mathbb I}\mbox{m} \left[(hh^\dagger)_{11}
  \lambda_{11}\lambda^*_{11}\right] =0$.  In this case, while the constraint
from the out-of-equilibrium condition is only slightly relaxed with respect to
the previous case ($|\lambda_{11}|^2 \lsim 10^4\,\xi\, M_{N_1}/M_P$), it
involves only the coupling $\lambda_{11}$.  It is then conceivable that some
mechanism could suppress just this particular entry, without affecting the
light neutrino mass scale (e.g.  a texture zero in the matrix of the
$\lambda$'s lifted by some higher order effect).  As in the case discussed in
section~\ref{sec:case1}, small values of $\lambda$ do not necessarily imply
that the $CP$ asymmetry is small, since the contribution in
eq.~(\ref{eq:cp-asymm-case3}) depends in a crucial way on the size of the
couplings $h$, and the factor of $|\lambda_{11}|^2$ in the denominator can
also enhance $\epsilon_{N_1\to F_1}$.  Also in this case, the scale of
leptogenesis could be much lower than the bound in eq.~(\ref{eq:M1limit0})
without being in conflict with other conditions.  For example, by taking
$|\lambda_{11}|$ of the order of the electron Yukawa coupling $\sim 10^{-6}$
the out-of-equilibrium condition can be satisfied for values of $M_{N_1}$ as
low as $\sim 1\,$TeV.  By assuming that the $\lambda$ couplings different from
$\lambda_{11}$ are at least of the order of the $\mu$ Yukawa $\sim 10^{-4}$,
and taking for the seesaw suppression factor $(h\,\sigma/M_F ) \sim 10^{-2}$,
we also obtain a reasonable value for the light neutrino mass scale.  At the
same time, the contribution to the $CP$ asymmetry in
eq.~(\ref{eq:cp-asymm-case3}) can remain as large as ${\cal O}(h^2)$.
%


\section{Conclusions}
\label{sec:conclusions}
In this paper we have explored the consequences for leptogenesis of adding to
the seesaw model a new scale related to the breaking of an additional
$U(1)_X$ symmetry.  Our framework is inspired by $U(1)$ models for flavor \`a
la Froggatt-Nielsen, but it differs from the usual schemes firstly in its
simplicity, and secondly because we do not impose particular conditions for
the values of the fundamental couplings, nor on the ratio between the symmetry
breaking scale and the masses of the heavy vectorlike fermions.  As a
consequence, while the model represents an interesting playground to study
variations of the standard leptogenesis scenario, it does not pretend to
account also for the pattern of fermion masses and mixings.  (This could still
be achieved in more complicated schemes in which $U(1)_X$ appears as a
component of the full flavor symmetry.)  We have found that in all the cases
in which leptogenesis occurs while the $U(1)_X$ symmetry is still unbroken,
the expressions for the $CP$ asymmetries in the decays of the lightest
Majorana neutrino $\epsilon_{N_1}$ differ from the standard case.  The most
interesting situations occur when the $N_1$ lifetime and the $CP$ asymmetry
$\epsilon_{N_1}$ are controlled by two different sets of couplings, and are
thus unrelated.  In these cases successful leptogenesis can be achieved even
at a scale as low as a few TeV, without conflicting with the requirement of
$N_1$ out-of-equilibrium decays, and ensuring at the same time a reasonable
value for the scale of the light neutrino masses.


\section*{Appendix A}
\label{sec:appendix}

In this section we present some analytic results for the minimal $2+2$ model
with only two right-handed neutrinos $N_{1,2}$ and two pairs of vectorlike
fermions $F_{1,2}$ and $\bar F_{1,2}$. This is the minimal field content that
ensures that two light neutrinos are massive, as is required by  neutrino
oscillations data. Since in this case one neutrino is exactly massless, the
neutrino mass squared differences measured in oscillations experiments
completely determine the light neutrinos masses, thus allowing for a certain
level of predictivity.  For a normal hierarchy spectrum we have $m_{\nu_2}
=\sqrt{\Delta m_{sol}^2}\approx 9\cdot 10^{-3}\,$eV and $m_{\nu_3} = \sqrt{\Delta
  m_{atm}^2+\Delta m_{sol}^2}\approx 5\cdot 10^{-2}\,$eV.  Focusing on the
case discussed in section~\ref{sec:case1}, that is defined by the condition
$M_{F}>M_{N}$, the $2\times 3$ matrix of the effective couplings
$\tilde{\lambda}$ can be written as \cite{Casas:2001sr}
%
\begin{equation}
\label{eq:tildelam}
\tilde{\lambda} = \frac{1}{v}\,M_N^{1/2}\, {\mathrm R}~ m_\nu^{1/2} U^{\dagger},
\end{equation}
where R is a complex $2\times 3$ matrix satisfying R$\,$R$^T =I$ and R$^T$R$\ 
=\ $diag$(0,1,1)$ \cite{aba06a-aba06b}, $U = U_D\cdot {\mathrm{diag}}(1, e^{-i
  \phi/2},1)$ with $U_D$ the leptonic mixing matrix, and $M_N$ and $m_\nu$ are
respectively the matrices of the heavy and light neutrinos mass eigenvalues.
Substituting (\ref{eq:tildelam}) in the combination of couplings that
determines the decay rate, eq.~(\ref{eq:width-decay-treel}), we obtain
\begin{equation}
\label{eq:tildelambdalambda11}
(\tilde{\lambda}\tilde{\lambda}^{\dagger})_{11} 
= \frac{M_{N_{1}}}{v^2} (m_{\nu_2} |R_{12}|^2 + m_{\nu_3} |R_{13}|^2).  
\end{equation}
Then eq.~(\ref{eq:out-of-eq}) implies the following constraint on the ratio of
the $U(1)_X$ breaking scale and the lightest Majorana mass $M_{N_1}$:
\begin{eqnarray}
\label{eq:sigmatoMn}
\xi \left( \frac{\sigma}{M_{N_{1}}}\right)^2  &\gsim& 10^{-5} 
\frac{M_P \sqrt{\Delta m_{atm}^2} }{v^2}\left[|R_{13}|^2+
\sqrt{\frac{\Delta m_{sol}^2}{ \Delta m_{atm}^2}}|R_{12}|^2 
\right]\\
&\approx& 
0.2\, \left(|R_{13}|^2+ 0.17\, |R_{12}|^2  \right).
\end{eqnarray}
This suggests that a mild hierarchy between $\sigma$ and $M_{N_1}$ is
certainly possible and, as was implicit in our scheme, allows for the
possibility that the three scales $M_F$, $M_{N_1}$ and $\sigma$ can
actually lie within a few orders of magnitude.

\section*{Acknowledgments}
We thank  Y. Nir for discussions and for pointing out an error in the first version of the paper. We thank  E. Roulet and S. Davidson for discussions.
D.A.S acknowledges M. Hirsch for his help
in some of the calculations. 
D.A.S is supported by a Spanish Ph.D. fellowship 
by M.C.Y.T. The work of M.L. is supported in 
part by Colciencias under contract number 
1115-333-18739. The work of E.N. is supported
in part by the Istituto Nazionale di Fisica 
Nucleare (INFN) in Italy, and by Colciencias in 
Colombia under contract 1115-333-18739.

\end{document}